\documentclass[10pt,prl,superscriptaddress,preprintnumbers,balancelastpage,twocolumn]{revtex4-2}
\usepackage{float}
\usepackage{graphicx}
\usepackage{epstopdf}
\usepackage{bm}
\usepackage[utf8]{inputenc}
\PassOptionsToPackage{hyphens}{url}
\PassOptionsToPackage{breaklinks=true}{hyperref}
\usepackage[hidelinks,colorlinks]{hyperref}
\usepackage{soul}
\usepackage{amsmath}
\usepackage{amssymb}
\usepackage{mathtools}
\usepackage{xfrac}
\usepackage{rotating}
\usepackage{enumitem}
\usepackage{scrextend}
\usepackage{slashed}
\usepackage{braket}
\usepackage{dcolumn}
\usepackage{caption}
\usepackage{subcaption}
\usepackage[dvipsnames]{xcolor}
\usepackage{tikz}
\usepackage{calligra}
\usetikzlibrary{calc}
\usetikzlibrary{arrows.meta, positioning}
\usepackage{url} 
\def\x{\boldsymbol{x}}
\def\s{\boldsymbol{s}}
\def\HR{\text{HR}}

\newcommand{\GlauberVertexJet}[1]{%
\begin{tikzpicture}[#1]%
\draw[fill, thick, GreenYellow] (0,0) circle (4pt); \draw[line width =1mm, olive] (0,0) circle (4pt);
\end{tikzpicture}%
}

\newcommand{\GlauberTextJet}{\raisebox{\dimexpr 2.0ex-\height-1pt\relax}{\GlauberVertexJet{}}}

\hypersetup{
    colorlinks=true,
    linkcolor=blue,
    citecolor=red,
    urlcolor=olive
}

\newcommand{\bea}{\begin{eqnarray}}
\newcommand{\eea}{\end{eqnarray}}
\newcommand{\beqa}{\begin{eqnarray}}
\newcommand{\eeqa}{\end{eqnarray}}
\newcommand{\be}{\begin{equation}}
\newcommand{\ee}{\end{equation}}
\newcommand{\beq}{\begin{equation}}
\newcommand{\eeq}{\end{equation}}

\newcommand{\Einan}[1]{\textcolor{red}{[#1]}}

\begin{document}

\preprint{USTC-ICTS/PCFT-26-46}

\title{
Spacelike-Collinear Scattering by the Method of Regions
}

\author{Wen Chen}
\email{chenwenphy@gmail.com}
\affiliation{School of Science and Engineering, The Chinese University of Hong Kong, Shenzhen, Longgang, 518172 Shenzhen, China}
\author{Einan Gardi}
\email{einan.gardi@ed.ac.uk}
\affiliation{Higgs Centre for Theoretical Physics, School of Physics and Astronomy, The University of Edinburgh, Edinburgh EH9 3FD, Scotland, UK}

\author{Rourou Ma}
\email{marr21@mail.ustc.edu.cn}
\affiliation{Interdisciplinary Center for Theoretical Study, \\ University of Science and Technology of China, Hefei, Anhui 230026, China}
\affiliation{Max-Planck-Institut f\"ur Physik, Werner-Heisenberg-Institut, Boltzmannstraße 8, 85748 Garching, Germany}

\author{Yao Ma}
\email{yaomay@phys.ethz.ch}
\affiliation{Institute for Theoretical Physics, ETH Zürich, 8093 Zürich, Switzerland}

\author{Yang Zhang}
\email{yzhphy@ustc.edu.cn}
\affiliation{Interdisciplinary Center for Theoretical Study, \\ University of Science and Technology of China, Hefei, Anhui 230026, China}
\affiliation{Peng Huanwu Center for Fundamental Theory, Hefei, Anhui 230026, China}
\affiliation{Center for High Energy Physics, Peking University,
Beijing 100871, People’s Republic of China}

\author{Zehao Zhu}
\email{zzhu3@ed.ac.uk}
\affiliation{Higgs Centre for Theoretical Physics, School of Physics and Astronomy, The University of Edinburgh, Edinburgh EH9 3FD, Scotland, UK}

\date{\today}

\begin{abstract}
We study the spacelike-collinear limit of gauge-theory scattering amplitudes using the Method of Regions. The corresponding splitting amplitude violates strict collinear factorisation through its dependence on the non-collinear partons. While the associated colour dependence has long been known, starting at two loops the splitting amplitude also acquires dependence on their kinematics. We show that this kinematic dependence originates from a unique hidden region present in the asymptotic expansion of the 5-point amplitude in the spacelike-collinear limit, but absent in the timelike limit. More generally, we propose that hidden regions provide the mechanism by which crossing-related asymptotic limits cease to be analytically connected.
We develop a general algorithm for the systematic identification of hidden regions. Applying it to the 5-point amplitude in super Yang--Mills theory, we compute the hidden-region contributions to the complete set of basis integrals and recover the exact kinematically dependent factorisation-violating splitting amplitude. In momentum space, the hidden region is characterised by soft and Glauber loop momenta. This explains why the Wilson-line calculation captures the complete kinematic dependence, thereby accounting for the observed universality across gauge theories.
\end{abstract}

\maketitle

A key property of on-shell gauge-theory scattering amplitudes is their factorisation in limits where some scattered partons become collinear~\cite{Bern:1994zx,Catani:1998nv,Catani:2003vu,Kosower:1999xi,Kosower:1999rx,Kosower:1999rx,Bern:2004cz,Badger:2004uk,Birthwright:2005ak,Catani:2011st,FshSmrSmk12,Badger:2015cxa,Schwartz:2017nmr,Becher:2019avh, Cieri:2024ytf,Duhr:2025lyg,Dixon:2019lnw,Henn:2024qjq,Buccioni:2026mfg,Duhr:2025cye,Duhr:2025lyg}. In such limits, the singular terms associated with the small scattering angle define a \emph{splitting amplitude} (${\mathbf{Sp}}$), which factorises from the rest of the process, an on-shell amplitude with fewer external legs.
Particularly, in the two-particle collinear limit $p_a\cdot p_b\to 0$, an $n$-parton amplitude admits~\cite{Kosower:1999xi}
\begin{align}
\label{AnCollFact}
\!{\cal A}_{n} (a^{\lambda_a},b^{\lambda_b},\cdots) 
\simeq 
\sum_{\lambda_P}{\mathbf{Sp}}_{\lambda_P}(a^{\lambda_a},b^{\lambda_b})
{\cal A}_{n-1}( P^{\lambda_{P}},\cdots),
\end{align}
where power corrections are neglected and the sum is over physical polarisations $\lambda_P$ of the parent parton $P$, whose momentum and colour are defined by summing those of the collinear partons $a$ and $b$. 
When the collinear partons are all in the final state (timelike splitting) this factorisation is strict: the splitting amplitude depends only on the subset of collinear particles. However, when initial- and final-state partons become collinear (spacelike splitting) strict factorisation is violated: the splitting amplitude acquires dependence on the rest of the process. 

While strict collinear factorisation violation (SCFV) due to colour correlations has been known for a while~\cite{Catani:2011st,FshSmrSmk12}, it was recently discovered that, from two loops, the splitting amplitude also depends on the kinematics of the rest of the process~\cite{Dixon:2019lnw,Henn:2024qjq,Buccioni:2026mfg}. 
Consequently, the timelike- and spacelike-collinear limits of the 5-point amplitude are not connected by analytic continuation, even though the corresponding physical regions are related by crossing~\cite{Bros:1985gy,Caron-Huot:2023ikn}.
Here we wish to explain why crossing-related asymptotic limits cease to be analytically connected.

The SCFV of two-loop splitting amplitudes cancels in the squared matrix elements upon summing over colour states~\cite{Henn:2024qjq,Buccioni:2026mfg}, and therefore does not affect inclusive cross sections at this order, in line with the factorisation theorems~\cite{Collins:1989gx,Collins:1985ue,Collins:1984kg}. 
A precise characterisation of the dynamical origin of amplitude-level SCFV is, however, a prerequisite for understanding how it manifests itself in less inclusive observables~\cite{Collins:2007nk,Rogers:2010dm,Rogers:2013zha} and jet cross sections~\cite{Forshaw:2006fk,Forshaw:2009sf,Forshaw:2012bi,Becher:2021zkk,Becher:2024kmk,Becher:2025igg,Becher:2026kbr,Dasgupta:2025cgl,Banfi:2025mra}, as well as for establishing an effective field-theory description of the collinear limit~\cite{Rothstein:2016bsq,Schwartz:2017nmr} and its generalisations~\cite{Cieri:2024ytf,Duhr:2025cye,Duhr:2025lyg}. 
  
The fundamental strategy to determine the behaviour of integrals and amplitudes in singular kinematic limits is the Method of Regions (MoR)~\cite{Beneke:1997zp,Smirnov:1998vk,Smirnov:1999bza,Smirnov:2002pj,Jantzen:2011nz,Semenova:2018cwy}. The key challenge in applying this method is to identify the complete set of regions~\cite{Gardi:2022khw,Ma:2023hrt,Ma25,Ma26}. While it is conceptually clear that all regions originate in Landau singularities~\cite{Landau:1959fi,Bjorken:1959fd,Nakanishi:1959jzx}, no general method exists to find them.
In momentum space this is a notoriously difficult problem, linked with characterisation of pinch surfaces.
In parameter space, for Euclidean integrals all singularities are associated with endpoint behaviour. In this special case a general tropical-geometry algorithm to find the set of regions has been established~\cite{Pak:2010pt}: regions correspond to certain facets of a Newton polytope. However, many physical limits of interest are inherently non-Euclidean, in which case pinch singularities can also occur in parameter space. Obtaining the asymptotic expansion of Feynman integrals by the MoR then requires additional \hbox{\emph{hidden regions}} (HR)~\cite{Jantzen:2012mw,Gardi:2024axt}, which are not captured by the facet construction. In this work, we propose an algorithm to find the complete set of HRs, and apply it to the 5-point amplitude in ${\cal N}=4$ super Yang-Mills (sYM) theory. 
We identify a unique HR which is responsible for the factorisation-violating dependence of the spacelike-collinear splitting amplitude on the kinematics of non-collinear partons. This exemplifies the general mechanism by which crossing-related asymptotic limits cease to be analytically connected due to HRs. 
Using parameter-space techniques~\cite{Chen:2019mqc,Chen:2019fzm,Chen:2020wsh,Chen:2025kdy,Chen:2023hmk} we directly compute the HR for the complete set of  basis integrals required to express the 5-point sYM amplitude, and recover the exact kinematically-dependent factorisation-violating splitting amplitude.

\section{5-point spacelike-collinear limit}

Given a 5-point massless gauge-theory scattering amplitude (\hbox{$p_i^2 = 0$}) for the process $1+2\to 3+4+5$,
we consider the expansion around the spacelike-collinear limit $2||3$ with the parametrisation~\cite{Henn:2024qjq},
\begin{eqnarray}
\label{param}
&&s_{12}= sz,\quad
s_{23}= -4 \delta^2, \quad
s_{34}=(1-z)xs, \quad
s_{45}= s,\nonumber \\
&&s_{15}=xs+
    \delta c,\qquad c\equiv
    \frac{8 y}{1+y^2}
    \sqrt\frac{s (-x)(1-x)z}{z-1} 
\end{eqnarray}
where~$\delta\to 0$.
The physical kinematic domain is~\hbox{$s>0$}  with   $-1<x<0$ and $z>1$. 
In these variables we describe the spacelike splitting amplitude as
\begin{equation}
{\mathbf{Sp}}: \qquad a^{\lambda_a}(zQ)\to P^{\lambda_P}\,(Q)
+b^{\lambda_b}((1-z)Q)\,,
\end{equation}
with $Q\simeq \sqrt{s}$.
Expanding (\ref{AnCollFact}) to two loops for $n=5$,
\begin{align}
\label{Sp2}
{\cal A}^{(2)}_{5} 
\simeq 
\sum_{\lambda_P}{\mathbf{Sp}}^{(0)}
{\cal A}^{(2)}_{4}
+
{\mathbf{Sp}}^{(1)}
{\cal A}^{(1)}_{4}
+
{\mathbf{Sp}}^{(2)}
{\cal A}^{(0)}_{4},
\end{align}
where ${\cal O}(\delta)$ power-suppressed terms are neglected. Here the $1+P\to 4+5$ amplitude ${\cal A}^{(0)}_{4}$ depends on $s$ and~$x$.
Strict collinear factorisation would require that 
${\mathbf{Sp}}^{(k)}$ only depends on $\delta$ and $z$. However, due to SCFV the splitting amplitude also depends on $x$ and $y$ from two-loop order. It may be expressed as~\cite{Dixon:2019lnw,Henn:2024qjq,Buccioni:2026mfg}:
\begin{equation}
\label{SpExp}
{\mathbf{Sp}}_{\lambda_P}(a^{\lambda_a},b^{\lambda_b})=ig_s {\text{Sp}}_{\lambda_P}(a^{\lambda_a},b^{\lambda_b}) \, e^{\bm{\mathcal{G}}_{\vec{\lambda}}(z;\epsilon)
    +{\bf \Delta}_{\vec{\lambda}}}
     {\bf R}_{\vec{\lambda}}(z)\,,
  \end{equation}  
where ${\text{Sp}}$ is the tree-level splitting amplitude with helicities 
$\vec{\lambda}\equiv \{\lambda_{P};\lambda_{a},\lambda_{b}\}$, and ${\bf R}_{\vec{\lambda}}(z) = {\bf{T}}_{ab}^c +{\cal }O(\alpha_s^2)$ with $a$, $b$ and $c$ the colour indices of the collinear particles $a$, $b$, and the parent~$P$, respectively. $\bm{\mathcal{G}}_{\vec{\lambda}}(z;\epsilon)$ starts at one loop and involves dependence on the colour of the non-collinear partons~\cite{Catani:2011st,Buccioni:2026mfg}, but its kinematic dependence is restricted to the collinear particles. Finally, ${\mathbf{\Delta}}_{\vec{\lambda}}$, the focus of this paper, starts at two loops:
\begin{align}
{\mathbf{\Delta}}_{\vec{\lambda}} = \left(\frac{\alpha_s (\mu^2)
}{4\pi}
\right)^2 \, {\mathbf{\Delta}}^{(2)}_{\vec{\lambda}}+\cdots .
\end{align}
${\mathbf{\Delta}}^{(2)}_{\vec{\lambda}}$ is universal -- it is the same in QCD and sYM~\cite{Buccioni:2026mfg} -- and features SCFV dependence on \emph{both the colour and the momenta} of non-collinear partons~\cite{Henn:2024qjq,Buccioni:2026mfg}:
\begin{align}
\label{Delta2}
    {\mathbf{\Delta}}_{\vec{\lambda}}^{(2)} \!= 4\pi i\Big(\frac{\mu^2\, z}{4\delta^2(z-1)}\Big)^{2\epsilon}\! 
    if^{adc} {\bf{T}}_1^a{\bf{T}}_3^c\!\!
    \sum_{I=4,5}\!\!
    {\bf{T}}_I^d \,g_{\vec{\lambda}}(z_I,\bar{z}_I;\epsilon),
\end{align}
with
\begin{eqnarray}
    g_{\vec{\lambda}}(z_I,\bar{z}_I;\epsilon)
   \!\! &=&\!\! \Big(\frac{1}{\epsilon^2}-\zeta_2\Big) (\ln |z_I|^2\!+\!2\pi i) +4\zeta_3 + \frac{h_{\vec{\lambda}}}{3} \gamma_I(y), \nonumber\\
   &&\hspace*{-60pt} \gamma_I(y)=\Big(\ln^2 \frac{z_I}{\bar{z}_I} +4\pi^2\Big) \ln \frac{z_I}{\bar{z}_I},\qquad z_I=\frac{\langle 23 \rangle \langle 1I \rangle}{\langle 12 \rangle \langle 3I \rangle}\,,
\end{eqnarray}
where $h_{\vec{\lambda}}=\lambda_b$ for $g\to gg$.
In terms of the parametrisation (\ref{param}):
\begin{align}
\label{z45_expressions}
&|z_4|^2=\frac{4\delta^2 (1+x)}{sz(z-1)(-x)},\qquad |z_5|^2=\frac{4\delta^2  (-x)}{sz(z-1)(1+x)},\nonumber\\
&\ln \frac{z_4}{\bar{z}_4} +i\pi 
=
\ln\frac{z_5}{\bar{z}_5} -i\pi 
=2\ln\left(\frac{i+y}{i-y}\right)
=
-4i \arctan (y)
\,.
\end{align}
Note that $\gamma_I(y)$ depend on the kinematics only through $\arctan(y)$, which is the dihedral angle between the ${\cal A}_4$ scattering plane and the splitting amplitude plane.  

The universality of ${\mathbf{\Delta}}^{(2)}_{\vec{\lambda}}$ and the fact that it is the only component of the splitting amplitude at this order to feature SCFV kinematic dependence make it very interesting. The dependence on the dihedral angle (on $y$) is the most striking factorisation-violating feature, as such dependence cannot be present in ${\cal A}_4$ or in the timelike splitting function. 
This motivates us to examine the asymptotic behaviour of ${\cal A}^{(2)}_{5}$ for $\delta\to 0$ in the spacelike domain, in order to explain why this expansion differs so radically from its timelike counterpart, and then recover Eq.~(\ref{Delta2}). We will work to leading order in $\delta$ and focus on terms scaling as~$\delta^{0-4\epsilon}$, which on general grounds cannot be attributed to the other terms in Eq.~(\ref{Sp2}).

\section{The Method of Regions}

The MoR expresses the asymptotic expansion of a multi-scale Feynman integral $I_\mathcal{G}$ as a sum over a number of ``region integrals'', $I_\mathcal{G}=\sum_R I_{\cal G}^R$, such that the expansion in region $R$ is performed subject to a distinct scaling law of the integration variables as the limit is approached, while the integration of $I_{\cal G}^R$ is carried over the entire (unrestricted) domain. 
When applying the MoR, the first and usually subtle step is to identify the complete set of regions. To this end, we apply the geometric approach~\cite{Pak:2010pt}. We express a Feynman integral $I_\mathcal{G}$ of graph $\mathcal{G}$ with $N$ propagators in the Lee-Pomeransky (LP) representation~\cite{Lee:2013hzt} where it is defined by a single polynomial $\mathcal{P}(\x,\s)=\mathcal{U}(\x)+\mathcal{F}(\x;\s)$, where $\mathcal{U}$ and $\mathcal{F}$ are the standard Symanzik graph polynomials, $\x$ represents the set of $N$ edge parameters $x_e$ and $\s$ the set of external kinematic variables. We associate $\mathcal{P}(\x,\s)$ with a Newton polytope $\Delta(\mathcal{G})$ in $\mathbb{R}^{N+1}$: each monomial 
\begin{equation}
\label{monomialDef}
m_i=c_i x_1^{r_{i,1}}\cdots x_N^{r_{i,N}}\in \mathcal{P} 
\quad \text{where} \ \ c_i\sim \mathcal{O}(\delta^{a_i})
\end{equation}
defines a point 
\begin{equation}
\vec{r}_i
\equiv
(\boldsymbol{r}_i;a_i)=(r_{i,1},\dots,r_{i,N};a_i)
\in \mathbb{R}^{N+1}\,.
\end{equation}
The convex hull of these points defines $\Delta(\mathcal{G})$.
Most regions correspond one-to-one to the \emph{lower facets} of $\Delta(\mathcal{G})$ (facets whose inward-pointing normal vectors have positive $(N+1)^\text{th}$ components), which we call \emph{facet regions}. With the normal vector defined as 
\begin{equation}
\vec{v}_R\equiv (\boldsymbol{v}_R;1) = (v_{R,1},\dots,v_{R,N};1)\,,
\end{equation}
the region $R$ is characterised by the parameter scaling law,
$
x_e\to x_e \delta^{v_{R,e}},
$ where each monomial (\ref{monomialDef})
acquires a \emph{weight}
$w_R(m_i)=\vec{v}_R\cdot {\vec{r}}_i$. The dominant monomials under this scaling law  
constitute the leading 
polynomial~$\mathcal{P}^{(R)}$, and the corresponding LP integral yields the leading power contribution to~$I_{\cal G}^R$.

In certain cases pinch singularities arise in the parametric representation due to cancellations among monomials of $\mathcal{F}$~\cite{Jantzen:2011nz,Gardi:2024axt}, giving rise to HRs that cannot be so simply identified. Such regions were discussed early on~\cite{Jantzen:2011nz} in the context of potential and Glauber physics. They could only be detected in special instances where the cancellation is realised linearly in~$x_e$.
Other examples of HRs in  $2\to 2$  processes with massless propagators have recently been studied~\cite{Gardi:2024axt,Ma25} both in wide-angle scattering and in the Regge limit. 
In general, HRs are characterised by a delicate interplay between scaling behaviour and cancellations: some monomials of~${\mathcal{F}}$, which we refer to as \emph{superleading}~(SL), are individually enhanced under the relevant scaling law~$x_e\to x_e \delta^{v_{\HR,e}}$ relative to all remaining monomials in $\mathcal{P}$, yet undergo mutual cancellations so that after applying the scaling, their sum, $\mathcal{F}_\text{SL}^{(\HR)}$, is of the same order as the remaining leading monomials. 
The central difficulty in identifying a HR is that the scaling law and the set of superleading monomials must be determined simultaneously.

\section{Hidden Region Finder} 

Our solution of the problem is based on the realisation that HRs must be associated with pinch singularities. In contrast to facet regions, HRs are not fully characterised by a scaling law of~$\{x_e\}$; additional relations between the parameters are necessary to reinforce cancellations in the superleading sector~\cite{Gardi:2024axt,Ma25}. These relations are the solution of the following Landau pinch conditions in parameter space
\begin{equation}
\label{Landau_HR}
\left\{\forall x_e\neq 0,\quad \left.
\frac{\partial \mathcal{F}_\text{SL}^{(\HR)}}{\partial x_e}=
\frac{\partial \mathcal{F}_\text{SL}}{\partial x_e}=0\right\}\right|_{ 
x_e>0; \,\,\s\in {\cal K}}\,,
\end{equation}
i.e., for a HR to exist the derivatives of the superleading polynomial $\mathcal{F}_\text{SL}$ with respect to all nonvanishing edge parameters must simultaneously vanish within the kinematic domain of interest $\s\in {\cal K}$ and for positive~$\{x_e\}$. The positivity condition is the usual requirement for a first sheet Landau singularity~\cite{Collins:2020euz}. 
We assume that the superleading monomials are all nonvanishing prior to applying the scaling,
i.e. they are part of $\mathcal{F}_0\equiv \lim_{\delta\to 0}\mathcal{F}$. Since the cancellation requires homogeneous weight for all the SL monomials,
\begin{align}
\label{vHR_homogen}
\begin{split}
w_{\HR}(m_i)&\equiv \vec{v}_{\text{HR}}\cdot {\vec{r}}_i=W_{\rm SL},
\qquad \forall m_i\in{\cal F}_{\rm SL}\,,
\end{split}
\end{align}
the pinch condition in Eq.~(\ref{Landau_HR}) equivalently holds either before or after applying the HR scaling.
Note that there is usually no pinch solution for the original ${\cal F}$ nor for $\mathcal{F}_0$, since $\mathcal{F}_0$ (usually)  contains additional monomials which do not participate in the cancellation; we call these the \emph{obstruction polynomial}.
A necessary (but insufficient) condition for a HR in a given topology is that the derivatives of \({\cal F}_0\) contain mixed-sign structures for $\{x_e>0\}$ within the kinematic domain $\s\in {\cal K}$.
If no such derivatives exist, the topology is discarded.
This requirement was proposed and used in~\cite{Gardi:2024axt} as a first selection criterion to identify integral topologies that potentially have~HRs.  
Once this first selection has been applied, we turn to the more refined algorithm, the Hidden Region Finder (HRF). Let us briefly summarise the key ideas, leaving the detailed exposition for a separate publication~\cite{HRF}.
The HRF consists of two main steps: (1) decomposing ${\cal F}_0$ into the SL sector and the obstruction, ${\cal F}_0=\mathcal{F}_\text{SL} + \mathcal{F}_\text{Obs}$, (2) solving for the HR scaling vector.

The decomposition stage begins by examining the derivatives of
\({\cal F}_0\).  A HR can only arise from derivative structures
which contain cancellations; accordingly we extract from these derivatives
the non-monomial mixed-sign polynomial factors $f_i$, denoted collectively by
$
{\cal C}=\{f_i\}$.
We then search for a subset of monomials of~\({\cal F}_0\), whose sum defines
\({\cal F}_{\rm Obs}\), such that
${\cal F}_{\rm SL}={\cal F}_0-{\cal F}_{\rm Obs}$
belongs to an ideal generated by products of candidate cancellation factors,
\begin{equation}
{\cal F}_{\rm SL}\in \langle g_1,\ldots,g_r\rangle,
\qquad
g_k(\x,\s)=\prod_{f_i\in S_k\subset{\cal C}} f_i(\x,\s)\,.
\end{equation}
Equivalently,
$
{\cal F}_{\rm SL}
=
\sum_k M_k(\x,\s)\,g_k(\x,\s)
$
for some polynomials \(M_k\).
Each generator~$g_k(\x,\s)$ is constructed as a product of  \emph{at least two} factors $f_i, f_j \in {\cal C}$ 
and its
monomial support must be compatible with monomials in \({\cal F}_0\).
A proposed generator must also satisfy the requirement  
that the factors $f_i$ it consists of have a
common positive solution in the relevant kinematic domain,
\begin{align}
\exists\,(\x,\s)\in \mathbb{R}_{>0}^{\,n}\times{\cal K}
\quad\text{such that }
f_i(\x,\s)=0
\ \ \forall \, f_i\in S_k .
\end{align}
After a candidate \({\cal F}_{\rm SL}\) has been identified, the same
condition is imposed on the union of all factors entering the generators that
actually appear in \({\cal F}_{\rm SL}\).  This realises the parameter-space
pinch condition~(\ref{Landau_HR}).

Having at hand an admissible decomposition of \({\cal F}_0\) into
\({\cal F}_{\rm SL}\) and \({\cal F}_{\rm Obs}\), we determine the
corresponding scaling law
\(\vec v_{\HR}=(\boldsymbol v_{\HR};1)\).  For each monomial
\(m_i\in\mathcal{P}\), with exponents \(\vec r_i=(\boldsymbol r_i;a_i)\), we use the
weight \(w_{\HR}(m_i)=\vec v_{\HR}\cdot\vec r_i\). We require homogeneity of the superleading sector (\ref{vHR_homogen}) and a hierarchy gap between the superleading monomials and all other monomials, 
\begin{align}
\label{vHR_hierarch}
\begin{split}
w_{\HR}(m_i)&\geq W_{\HR}>W_{\rm SL},
\qquad 
\forall m_i\in\mathcal{P}\setminus{\cal F}_{\rm SL}\,,
\end{split}
\end{align}
where \(W_{\HR}\) is the weight of the dominant monomials outside the superleading set. We conjecture that, for any admissible decomposition corresponding to a HR, the homogeneity (\ref{vHR_homogen}) and hierarchy (\ref{vHR_hierarch}) conditions uniquely determine $\vec{v}_{\HR}$. Indeed, any residual continuous freedom in the scaling would induce an additional rescaling symmetry of the region-expanded integrand and hence a scaleless integral. The hierarchy gap $W_{\HR}-W_{\rm SL}>0$ is then fixed uniquely and measures the depth of the cancellation.

\section{MoR application to 5-point integrals}

We now apply the MoR to the two-loop 5-point sYM amplitude
\({\cal A}^{(2)}_{5}\) in the spacelike-collinear limit
\(\delta\to0\) defined in Eq.~(\ref{param}).
The discussion above suggests a useful diagnostic.  Facet regions are determined by the Newton polytope of the LP polynomial and therefore do not depend on the signs of the kinematic invariants; they are present in both the spacelike- and timelike-collinear limits.
HRs, by contrast, rely on sign-dependent cancellations in the leading polynomial. It follows that the terms in Eq.~(\ref{Delta2}) responsible for SCFV in the spacelike limit, which have no counterpart in the timelike case, must be associated with HRs rather than facet regions.

To explore this we express the
amplitude in terms of a basis of uniformly transcendental (UT) integrals. As a preliminary step, following~\cite{Henn:2024qjq}, we obtain these integrals as a solution of canonical differential equations and verify that they reproduce the  asymptotic expansion of ${\cal A}_5$ at leading order in $\delta$. 

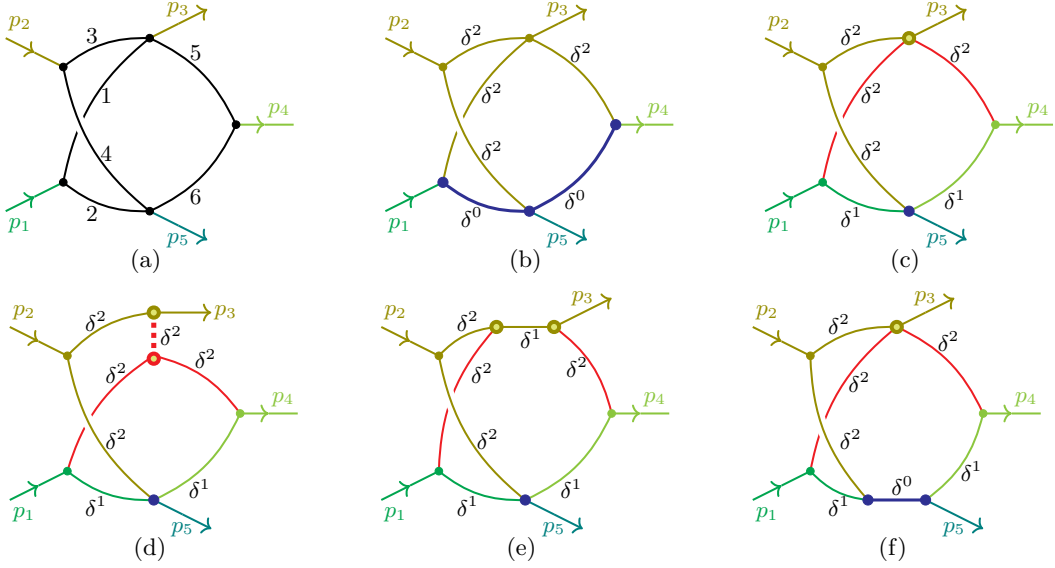
\begin{figure*}[htb]
\centering
\begin{subfigure}[b]{0.25\textwidth}
\centering
\resizebox{0.9\textwidth}{!}{
\begin{tikzpicture}[baseline=11ex, scale=0.4]

\path (5,8) edge [thick, bend right = 20] (2,3) {};
\path (5,8) edge [thick, bend right = 20] (2,7) {};
\draw (5,2) edge [thick, draw=white, double=white, double distance=3pt, bend left = 20] (2,7) node [] {};\draw (5,2) edge [thick, bend left = 20] (2,7) node [] {};
\path (5,2) edge [thick, bend left = 20] (2,3) {};
\path (5,8) edge [thick, bend left = 20] (8,5) {};
\path (5,2) edge [thick, bend right = 20] (8,5) {};

\path (0,2) edge [->, thick, Green] (1,2.5) {};\path (1,2.5) edge [thick, Green] (2,3) {};
\path (0,8) edge [->, thick, olive] (1,7.5) {};\path (1,7.5) edge [thick, olive] (2,7) {};
\path (8,5) edge [->, thick, LimeGreen] (9,5) {};\path (9,5) edge [thick, LimeGreen] (10,5) {};
\path (7,9) edge [<-, thick, olive] (6,8.5) {};\path (6,8.5) edge [thick, olive] (5,8) {};
\path (7,1) edge [<-, thick, teal] (6,1.5) {};\path (6,1.5) edge [thick, teal] (5,2) {};

\node [draw, circle, fill, minimum size=3pt, inner sep=0pt, outer sep=0pt] () at (5,8) {};
\node [draw, circle, fill, minimum size=3pt, inner sep=0pt, outer sep=0pt] () at (2,3) {};
\node [draw, circle, fill, minimum size=3pt, inner sep=0pt, outer sep=0pt] () at (5,2) {};
\node [draw, circle, fill, minimum size=3pt, inner sep=0pt, outer sep=0pt] () at (2,7) {};
\node [draw, circle, fill, minimum size=3pt, inner sep=0pt, outer sep=0pt] () at (8,5) {};

\node () at (0.5,1.5) {\color{Green} $p_1$};
\node () at (0.5,8.5) {\color{olive} $p_2$};
\node () at (9.5,5.5) {\color{LimeGreen} $p_4$};
\node () at (6,1) {\color{teal} $p_5$};
\node () at (6,9) {\color{olive} $p_3$};

\node () at (3.5,6) {\small $1$};
\node () at (3,1.9) {\small $2$};
\node () at (3,8.1) {\small $3$};
\node () at (3.5,4) {\small $4$};
\node () at (6.6,7.5) {\small $5$};
\node () at (6.6,2.5) {\small $6$};

\end{tikzpicture}}
\vspace{-1em}
\caption{}
\label{cdep_graph_seed_topology}
\end{subfigure}
\hspace{1em}
\begin{subfigure}[b]{0.25\textwidth}
\centering
\resizebox{0.9\textwidth}{!}{
\begin{tikzpicture}[baseline=11ex, scale=0.4]

\path (5,8) edge [thick, olive, bend right = 20] (2,3) {};
\path (5,8) edge [thick, olive, bend right = 20] (2,7) {};
\draw (5,2) edge [thick, draw=white, double=white, double distance=3pt, bend left = 20] (2,7) node [] {};\draw (5,2) edge [thick, olive, bend left = 20] (2,7) node [] {};
\path (5,2) edge [very thick, Blue, bend left = 20] (2,3) {};
\path (5,8) edge [thick, olive, bend left = 20] (8,5) {};
\path (5,2) edge [very thick, Blue, bend right = 20] (8,5) {};

\path (0,2) edge [->, thick, Green] (1,2.5) {};\path (1,2.5) edge [thick, Green] (2,3) {};
\path (0,8) edge [->, thick, olive] (1,7.5) {};\path (1,7.5) edge [thick, olive] (2,7) {};
\path (8,5) edge [->, thick, LimeGreen] (9,5) {};\path (9,5) edge [thick, LimeGreen] (10,5) {};
\path (7,9) edge [<-, thick, olive] (6,8.5) {};\path (6,8.5) edge [thick, olive] (5,8) {};
\path (7,1) edge [<-, thick, teal] (6,1.5) {};\path (6,1.5) edge [thick, teal] (5,2) {};

\node [draw, circle, minimum size=3pt, color=olive, fill=olive, inner sep=0pt, outer sep=0pt] () at (5,8) {};
\node [draw, circle, minimum size=4pt, color=Blue, fill=Blue, inner sep=0pt, outer sep=0pt] () at (5,2) {};
\node [draw, circle, minimum size=4pt, color=Blue, fill=Blue, inner sep=0pt, outer sep=0pt] () at (2,3) {};
\node [draw, circle, minimum size=3pt, color=olive, fill=olive, inner sep=0pt, outer sep=0pt] () at (2,7) {};
\node [draw, circle, minimum size=4pt, color=Blue, fill=Blue, inner sep=0pt, outer sep=0pt] () at (8,5) {};

\node () at (0.5,1.5) {\color{Green} $p_1$};
\node () at (0.5,8.5) {\color{olive} $p_2$};
\node () at (9.5,5.5) {\color{LimeGreen} $p_4$};
\node () at (6,1) {\color{teal} $p_5$};
\node () at (6,9) {\color{olive} $p_3$};

\node () at (3.7,6.1) {\small $\delta^{2}$};
\node () at (3,1.8) {\small $\delta^{0}$};
\node () at (3,8.1) {\small $\delta^{2}$};
\node () at (3.7,4.1) {\small $\delta^{2}$};
\node () at (6.8,7.5) {\small $\delta^{2}$};
\node () at (6.6,2.4) {\small $\delta^{0}$};
\end{tikzpicture}}
\vspace{-1em}
\caption{}
\label{cdep_graph_subtopology_collinear_region}
\end{subfigure}
\hspace{1em}
\begin{subfigure}[b]{0.25\textwidth}
\centering
\resizebox{0.9\textwidth}{!}{
\begin{tikzpicture}[baseline=11ex, scale=0.4]

\path (5,8) edge [thick, Red, bend right = 20] (2,3) {};
\path (5,8) edge [thick, olive, bend right = 20] (2,7) {};
\draw (5,2) edge [thick, draw=white, double=white, double distance=3pt, bend left = 20] (2,7) node [] {};\draw (5,2) edge [thick, olive, bend left = 20] (2,7) node [] {};
\path (5,2) edge [thick, Green, bend left = 20] (2,3) {};
\path (5,8) edge [thick, Red, bend left = 20] (8,5) {};
\path (5,2) edge [thick, LimeGreen, bend right = 20] (8,5) {};

\path (0,2) edge [->, thick, Green] (1,2.5) {};\path (1,2.5) edge [thick, Green] (2,3) {};
\path (0,8) edge [->, thick, olive] (1,7.5) {};\path (1,7.5) edge [thick, olive] (2,7) {};
\path (8,5) edge [->, thick, LimeGreen] (9,5) {};\path (9,5) edge [thick, LimeGreen] (10,5) {};
\path (7,9) edge [<-, thick, olive] (6,8.5) {};\path (6,8.5) edge [thick, olive] (5,8) {};
\path (7,1) edge [<-, thick, teal] (6,1.5) {};\path (6,1.5) edge [thick, teal] (5,2) {};

\draw[fill, thick, White] (5,8) circle (5pt); \draw (5,7.8) edge [line width = 1mm, GreenYellow] (5,8.2) node [] {}; \draw[ultra thick, olive] (5,8) circle (5pt);
\node [draw, circle, minimum size=4pt, color=Blue, fill=Blue, inner sep=0pt, outer sep=0pt] () at (5,2) {};
\node [draw, circle, minimum size=3pt, color=Green, fill=Green, inner sep=0pt, outer sep=0pt] () at (2,3) {};
\node [draw, circle, minimum size=3pt, color=olive, fill=olive, inner sep=0pt, outer sep=0pt] () at (2,7) {};
\node [draw, circle, minimum size=3pt, color=LimeGreen, fill=LimeGreen, inner sep=0pt, outer sep=0pt] () at (8,5) {};

\node () at (0.5,1.5) {\color{Green} $p_1$};
\node () at (0.5,8.5) {\color{olive} $p_2$};
\node () at (9.5,5.5) {\color{LimeGreen} $p_4$};
\node () at (6,1) {\color{teal} $p_5$};
\node () at (6,9) {\color{olive} $p_3$};

\node () at (3.7,6.1) {\small $\delta^{2}$};
\node () at (3,1.8) {\small $\delta^{1}$};
\node () at (3,8.1) {\small $\delta^{2}$};
\node () at (3.7,4.1) {\small $\delta^{2}$};
\node () at (6.8,7.5) {\small $\delta^{2}$};
\node () at (6.6,2.4) {\small $\delta^{1}$};
\end{tikzpicture}}
\vspace{-1em}
\caption{}
\label{cdep_graph_subtopology_Glauber_region}
\end{subfigure}
\\
\begin{subfigure}[b]{0.25\textwidth}
\centering
\resizebox{0.9\textwidth}{!}{
\begin{tikzpicture}[baseline=11ex, scale=0.4]

\path (5,7) edge [thick, Red, bend right = 20] (2,3) {};
\path (5,8.5) edge [thick, olive, bend right = 20] (2,7) {};
\path (5,8.5) edge [line width=2pt, Red, dotted] (5,7) {};
\draw (5,2) edge [thick, draw=white, double=white, double distance=3pt, bend left = 20] (2,7) node [] {};\draw (5,2) edge [thick, olive, bend left = 20] (2,7) node [] {};
\path (5,2) edge [thick, Green, bend left = 20] (2,3) {};
\path (5,7) edge [thick, Red, bend left = 20] (8,5) {};
\path (5,2) edge [thick, LimeGreen, bend right = 20] (8,5) {};

\path (0,2) edge [->, thick, Green] (1,2.5) {};\path (1,2.5) edge [thick, Green] (2,3) {};
\path (0,8) edge [->, thick, olive] (1,7.5) {};\path (1,7.5) edge [thick, olive] (2,7) {};
\path (8,5) edge [->, thick, LimeGreen] (9,5) {};\path (9,5) edge [thick, LimeGreen] (10,5) {};
\path (7,8.5) edge [<-, thick, olive] (5,8.5) {};
\path (7,1) edge [<-, thick, teal] (6,1.5) {};\path (6,1.5) edge [thick, teal] (5,2) {};

\draw[fill, thick, White] (5,8.5) circle (5pt); \draw (5,8.3) edge [line width = 1mm, GreenYellow] (5,8.7) node [] {}; \draw[ultra thick, olive] (5,8.5) circle (5pt);
\draw[fill, thick, White] (5,6.9) circle (5pt); \draw (5,6.7) edge [line width = 1mm, GreenYellow] (5,7.1) node [] {}; \draw[ultra thick, Red] (5,6.9) circle (5pt);
\node [draw, circle, fill=Green, Green, minimum size=3pt, inner sep=0pt, outer sep=0pt] () at (2,3) {};
\node [draw, circle, fill=Blue, Blue, minimum size=4pt, inner sep=0pt, outer sep=0pt] () at (5,2) {};
\node [draw, circle, fill=olive, olive, minimum size=3pt, inner sep=0pt, outer sep=0pt] () at (2,7) {};
\node [draw, circle, fill=LimeGreen, LimeGreen, minimum size=3pt, inner sep=0pt, outer sep=0pt] () at (8,5) {};

\node () at (0.5,1.5) {\color{Green} $p_1$};
\node () at (0.5,8.5) {\color{olive} $p_2$};
\node () at (9.5,5.5) {\color{LimeGreen} $p_4$};
\node () at (6,1) {\color{teal} $p_5$};
\node () at (7.5,8.5) {\color{olive} $p_3$};

\node () at (3.7,6.4) {\small $\delta^{2}$};
\node () at (3,1.8) {\small $\delta^{1}$};
\node () at (3,8.2) {\small $\delta^{2}$};
\node () at (3.7,4.1) {\small $\delta^{2}$};
\node () at (5.6,7.7) {\small $\delta^{2}$};
\node () at (6.8,6.9) {\small $\delta^{2}$};
\node () at (6.6,2.4) {\small $\delta^{1}$};
\end{tikzpicture}}
\vspace{-1em}\caption{}
\label{possibly_hidden_singular_graph_t_dot}
\end{subfigure}
\quad
\centering
\begin{subfigure}[b]{0.25\textwidth}
\centering
\resizebox{0.9\textwidth}{!}{
\begin{tikzpicture}[baseline=11ex, scale=0.4]

\path (4,8) edge [thick, Red, bend right = 20] (2,3) {};
\path (4,8) edge [thick, olive, bend right = 20] (2,7) {};
\path (4,8) edge [thick, olive] (6,8) {};
\draw (5,2) edge [thick, draw=white, double=white, double distance=3pt, bend left = 20] (2,7) node [] {};\draw (5,2) edge [thick, olive, bend left = 20] (2,7) node [] {};
\path (5,2) edge [thick, Green, bend left = 20] (2,3) {};
\path (6,8) edge [thick, Red, bend left = 20] (8,5) {};
\path (5,2) edge [thick, LimeGreen, bend right = 20] (8,5) {};

\path (0,2) edge [->, thick, Green] (1,2.5) {};\path (1,2.5) edge [thick, Green] (2,3) {};
\path (0,8) edge [->, thick, olive] (1,7.5) {};\path (1,7.5) edge [thick, olive] (2,7) {};
\path (8,5) edge [->, thick, LimeGreen] (9,5) {};\path (9,5) edge [thick, LimeGreen] (10,5) {};
\path (8,9) edge [<-, thick, olive] (6,8) {};
\path (7,1) edge [<-, thick, teal] (6,1.5) {};\path (6,1.5) edge [thick, teal] (5,2) {};

\draw[fill, thick, White] (4,8) circle (5pt); \draw (4,7.8) edge [line width = 1mm, GreenYellow] (4,8.2) node [] {}; \draw[ultra thick, olive] (4,8) circle (5pt);
\draw[fill, thick, White] (6,8) circle (5pt); \draw (6,7.8) edge [line width = 1mm, GreenYellow] (6,8.2) node [] {}; \draw[ultra thick, olive] (6,8) circle (5pt);
\node [draw, circle, fill=Green, Green, minimum size=3pt, inner sep=0pt, outer sep=0pt] () at (2,3) {};
\node [draw, circle, fill=Blue, Blue, minimum size=4pt, inner sep=0pt, outer sep=0pt] () at (5,2) {};
\node [draw, circle, fill=olive, olive, minimum size=3pt, inner sep=0pt, outer sep=0pt] () at (2,7) {};
\node [draw, circle, fill=LimeGreen, LimeGreen, minimum size=3pt, inner sep=0pt, outer sep=0pt] () at (8,5) {};

\node () at (0.5,1.5) {\color{Green} $p_1$};
\node () at (0.5,8.5) {\color{olive} $p_2$};
\node () at (9.5,5.5) {\color{LimeGreen} $p_4$};
\node () at (6,1) {\color{teal} $p_5$};
\node () at (7,9) {\color{olive} $p_3$};

\node () at (3.5,6.4) {\small $\delta^{2}$};
\node () at (3,1.8) {\small $\delta^{1}$};
\node () at (3,8.2) {\small $\delta^{2}$};
\node () at (3.7,4.1) {\small $\delta^{2}$};
\node () at (5.2,7.6) {\small $\delta^{1}$};
\node () at (6.8,6.6) {\small $\delta^{2}$};
\node () at (6.6,2.4) {\small $\delta^{1}$};
\end{tikzpicture}}
\vspace{-1em}\caption{}
\label{possibly_hidden_singular_graph_s_dot}
\end{subfigure}
\quad
\begin{subfigure}[b]{0.25\textwidth}
\centering
\resizebox{0.9\textwidth}{!}{
\begin{tikzpicture}[baseline=11ex, scale=0.4]

\path (5,8) edge [thick, Red, bend right = 20] (2,3) {};
\path (5,8) edge [thick, olive, bend right = 20] (2,7) {};
\draw (4,2) edge [thick, draw=white, double=white, double distance=3pt, bend left = 20] (2,7) node [] {};\draw (4,2) edge [thick, olive, bend left = 20] (2,7) node [] {};
\path (4,2) edge [thick, Green, bend left = 20] (2,3) {};
\path (5,8) edge [thick, Red, bend left = 20] (8,5) {};
\path (6,2) edge [thick, LimeGreen, bend right = 20] (8,5) {};
\path (6,2) edge [very thick, Blue] (4,2) {};

\path (0,2) edge [->, thick, Green] (1,2.5) {};\path (1,2.5) edge [thick, Green] (2,3) {};
\path (0,8) edge [->, thick, olive] (1,7.5) {};\path (1,7.5) edge [thick, olive] (2,7) {};
\path (8,5) edge [->, thick, LimeGreen] (9,5) {};\path (9,5) edge [thick, LimeGreen] (10,5) {};
\path (7,9) edge [<-, thick, olive] (6,8.5) {};\path (6,8.5) edge [thick, olive] (5,8) {};
\path (8,1) edge [<-, thick, teal] (6,2) {};

\draw[fill, thick, White] (5,8) circle (5pt); \draw (5,7.8) edge [line width = 1mm, GreenYellow] (5,8.2) node [] {}; \draw[ultra thick, olive] (5,8) circle (5pt);
\node [draw, circle, fill=Green, Green, minimum size=3pt, inner sep=0pt, outer sep=0pt] () at (2,3) {};
\node [draw, circle, fill=Blue, Blue, minimum size=4pt, inner sep=0pt, outer sep=0pt] () at (4,2) {};
\node [draw, circle, fill=Blue, Blue, minimum size=4pt, inner sep=0pt, outer sep=0pt] () at (6,2) {};
\node [draw, circle, fill=olive, olive, minimum size=3pt, inner sep=0pt, outer sep=0pt] () at (2,7) {};
\node [draw, circle, fill=LimeGreen, LimeGreen, minimum size=3pt, inner sep=0pt, outer sep=0pt] () at (8,5) {};

\node () at (0.5,1.5) {\color{Green} $p_1$};
\node () at (0.5,8.5) {\color{olive} $p_2$};
\node () at (9.5,5.5) {\color{LimeGreen} $p_4$};
\node () at (7,1) {\color{teal} $p_5$};
\node () at (6,9) {\color{olive} $p_3$};

\node () at (3.7,6.1) {\small $\delta^{2}$};
\node () at (3,1.8) {\small $\delta^{1}$};
\node () at (3,8.1) {\small $\delta^{2}$};
\node () at (3.4,4.1) {\small $\delta^{2}$};
\node () at (6.8,7.5) {\small $\delta^{2}$};
\node () at (5.2,2.5) {\small $\delta^{0}$};
\node () at (7.6,3) {\small $\delta^{1}$};
\end{tikzpicture}}
\vspace{-1em}\caption{}
\label{possibly_hidden_singular_graph_dot_s}
\end{subfigure}
\vspace{-1em}\caption{The seed topology (a) and its collinear (facet) region (b) and Glauber (hidden) region (c). (d)--(f): some HRs of other two-loop graphs, all of which arise from expanding the four-point vertices of (c). Four shades of green depict collinear momentum scaling along $p_1, p_2, p_4$ and $p_5$ (with possibly different virtualities) while red edges depict soft scaling as in Eq.~(\ref{eq:MomScal}). For each region we have indicated the scaling of line-momentum virtuality.}
\label{figure-cdep_graph_subtopology_representative_regions}
\end{figure*}

Our next task, following the above diagnostic, is to identify which integrals in this UT basis have HRs. To this end we first classify the scalar topologies appearing in each of these integrals.
This yields 469 scalar graphs.  
The mixed-sign derivative preselection~\cite{Gardi:2024axt}, which is only a necessary condition for a HR, selects 143 candidates.  Since $p_4$ and $p_5$ play no distinct role in the spacelike-collinear limit considered here, we identify graphs related by this permutation, leaving 70 representative topologies.
Next, applying the HRF algorithm, all planar topologies are discarded, while among the 36 nonplanar ones, double-pentagon and hexagon-box topologies, HR are found in just 16 representatives. These are organised around a single six-propagator seed topology, shown in Fig.~\ref{cdep_graph_seed_topology}.  As an unlabelled graph, the seed is a contraction minor of all 36 nonplanar representatives. What distinguishes the 16 HR-positive cases is instead how the external legs are attached: in each of them the contraction exposes the seed with the required external labelling. 
A reassuring observation is that the UT integrals which feature the 16 HR-positive topologies (or their $p_4 \leftrightarrow p_5$ permutations) are precisely those that feature $y$ dependence, which distinguishes the asymptotic expansion in the spacelike region from that in the timelike one. 

Let us now examine the seed integral in more detail. For the computation it is convenient to write it in projective space by introducing an additional parameter $x_0$:
\begin{align}
\label{IsixProp}
&I_{\mathcal{G}_{\bullet\bullet}}=\int_0^\infty \! \!
 {\cal D}{\bf x}\,{\cal I}_{\mathcal{G}_{\bullet\bullet}}
=\int_0^\infty\!\!\!
\frac{{\cal D}{\bf x}\, \Gamma_D}{
x_0^{7-\frac{3}{2}D}\left({\cal F}_0 
    +{\cal F}_\delta +x_0 {\cal U}-i\varepsilon\right)^{\frac{D}{2}} }\nonumber
\\
&{\cal D}{\bf x}\equiv 
\mathrm{d}x_0\mathrm{d}x_1\mathrm{d}x_2\cdots\mathrm{d}x_{6}\, \delta(1-\mathcal{E}({\bf x}))\,,
\end{align}
where $\Gamma_D
\equiv\frac{\Gamma(D/2)}{\Gamma(3D/2-6)}$
and~$\mathcal{E}({\bf x})$ is a degree-1 homogeneous function of $\{x_i\}$ and the graph polynomials are:
\begin{align}
{\cal F}_0
&=
\underbrace{
{\color{red}s \left(x_{2,5}+x x_{1,6}\right)
\left[(z-1)x_3-x_4\right]}
}_{{\cal F}_{\rm SL}}
+
\underbrace{\color{blue}
s z (x+1) x_{2,3,6}
}_{{\cal F}_{\rm Obs}}\, ,
\label{eq:fundamental_topology_superleadingFPlusObs}
\\
{\cal F}_\delta   &= {\color{blue}- c\delta  x_{1,4,6}}
+ 
c\delta  x_{2,3,6}  
  {  \color{blue}
\,+ \, 
4\delta^2 x_{1,4,5}}
- 
4\delta^2 x_{2,3,5}\,,\nonumber
\\
 {\cal U}   &= {\color{blue}x_{1,3}+x_{1,4}+x_{1,5}}+x_{1,6}+x_{2,3}+x_{2,4}\nonumber\\
&\quad+x_{2,5}+x_{2,6}+{\color{blue}x_{3,5}}+x_{3,6}+{\color{blue}x_{4,5}}+x_{4,6}\,,\nonumber
\end{align}
where $x_{i,j,\dots}$ is short for $x_ix_j\cdots$.

Beyond the hard region, $I_{\mathcal{G}_{\bullet\bullet}}$ has a  single facet region scaling as~\(\delta^{-4\epsilon}\) with the region vector 
\begin{equation}
\label{vFacetRegion}
\vec{v}_{C}=
(v_{1},v_{2},v_{3},v_{4},v_{5},v_{6};1)\!=\!
(-2, 0, -2, -2, -2, 0;1)\,.
\end{equation}
Since this region is present in both the spacelike and timelike domains, it cannot account for SCFV. Indeed, applying the facet region scaling~(\ref{vFacetRegion}) to $\mathcal{P}_{\mathcal{G}_{\bullet\bullet}}$ the leading power is independent of~$c$ and the dependence on $x$ can be scaled out sending $x_6\to x_6/x$.  

Let us now focus on the HR of~$I_{\mathcal{G}_{\bullet\bullet}}$.
The mixed-sign derivatives of \({\cal F}_0\) pass the HRF preselection. Despite this, it does not have a positive Landau solution prior to scaling. The HRF identifies one generator, yielding the decomposition of ${\cal F}_0$ in Eq.~(\ref{eq:fundamental_topology_superleadingFPlusObs}): the terms in red form the superleading sector \({\cal F}_{\rm SL}\), while the remaining
monomial \({\color{blue}s z(1+x)x_{2,3,6}}\) is the obstruction. Solving the conditions~(\ref{vHR_homogen}) 
and~(\ref{vHR_hierarch})
for the scaling vector $x_i\to x_i \delta^{v_i}$ we get 
\begin{equation}
\label{vHR_sol}
\vec{v}_{\HR}
\!= \!(v_{1},v_{2},v_{3},v_{4},v_{5},v_{6};\!1)\!=\!
(-2, -1, -2, -2, -2, -1;\!1)
\end{equation}
with which one can readily verify that the SL monomials scale as $\delta^{W_{\text{SL}}}=\delta^{-5}$ while the obstruction scales as $\delta^{W_{\text{HR}}}=\delta^{-4}$, similarly to the leading terms in the~${\cal F}_\delta$ and~$ {\cal U}$ polynomials (marked in blue in Eq.~(\ref{eq:fundamental_topology_superleadingFPlusObs})). Importantly, a term involving $c\delta$ is part of the leading~${\cal O}(\delta^{-4})$ HR polynomial, bringing in $y$ dependence at leading power.
Finally, let us examine the analytic structure of the HR integral. While locally the causal $-i\varepsilon$ instructs us how to deform the contour, it does not translate into global assignment of imaginary parts to $x$ and $z$. The HR integral needs to be dissected~\cite{Gardi:2024axt,Jones:2025jzc} along the hypersurfaces defined by ${\cal F}_{\rm SL}=0$, in order to obtain integrals that can be analytically continued.  
We conclude that in contrast to the facet region of~$I_{\mathcal{G}_{\bullet\bullet}}$ which is independent of $x$ and~$y$, the HR depends on both. 

Next let us interpret the regions in momentum space. To this end we use momentum conservation and the relation between the scaling of the edge parameters $x_e$ and the virtualities, $x_e\sim (q_e^2)^{-1}$, where $q_e$ is the edge momentum~\cite{Engel:2022kde,Gardi:2022khw}.
Fig.~\ref{cdep_graph_seed_topology} features two facet regions: the hard region  where both loop momenta $\sim\sqrt{s}$, and the collinear region (Fig.~\ref{cdep_graph_subtopology_collinear_region}) where both loop momenta are in the collinear-$23$ mode, $k^\mu = (k^+,k^-,\boldsymbol{k}_\perp) \sim \sqrt{s}\,(1,\delta^2,\delta)$, with $p_1$ and $p_2$ the lightcone ``$-$'' and ``$+$''  directions, respectively. The only HR features a soft loop and a Glauber loop (Fig.~\ref{cdep_graph_subtopology_Glauber_region}), with momenta
\begin{align}
\label{eq:MomScal}
    k_S\sim \sqrt{s}\,(\delta, \delta, \delta),\qquad k_G\sim \sqrt{s}\,(\delta, \delta^2, \delta),
\end{align}
respectively, where $k_S$ flows through the propagators $\{1,2,5,6\}$ while $k_G$ through $\{1,2,3,4\}$ or $\{3,4,5,6\}$ depending on the routing. The symbol \GlauberTextJet\ indicates the $23$-jet vertex where a Glauber momentum is exchanged. We emphasise that this HR features a Glauber-mode loop without Glauber propagators.
The loop momenta~(\ref{eq:MomScal}) can also be shown~\cite{LongerPaper} (see also~\cite{Nabeebaccus:2024mia}) to follow directly from the pinch conditions. 

Turning now to the complete set of $2\times 16$ graphs containing the HR, we find that the corresponding momentum flows remain compatible with Eq.~(\ref{eq:MomScal}), all inherited from the seed topology in Fig.~\ref{cdep_graph_subtopology_Glauber_region} by expanding its four-vertices into pairs of three-vertices. For example, a Glauber propagator ({\color{Red}\textbf{dotted}} line) arises if the upper four-vertex is expanded as in Fig.~\ref{possibly_hidden_singular_graph_t_dot}, a collinear-$23$ propagator (with momentum virtuality $\delta^{1}$) arises if this vertex is expanded otherwise (Fig.~\ref{possibly_hidden_singular_graph_s_dot}), and a hard propagator arises if the lower four-vertex is expanded (Fig.~\ref{possibly_hidden_singular_graph_dot_s}). 

The computation of the HR integrals corresponding to the UT basis is briefly described in ``End Matter'' part~{\bf A}. It will be discussed in detail in~\cite{LongerPaper}. The results are provided in~\cite{MathematicaNotebookTwoLoop} where one can verify that the transcendental functions of $x$ and $y$ in the HR have the expected uniformly maximal weight  (weight 4). 
Next we assemble the amplitude picking only the HR for each UT integral. This is done first in the trace basis~\cite{Henn:2024qjq} and then in a suitable colour-flow basis~\cite{Abreu:2024xoh} originally used for the multi-Regge limit. This is briefly described in
``End Matter'' part~{\bf B} and the results for the HR of the asymptotic  amplitude ${\cal A}^{(2)}_{5}$ are provided  in~\cite{MathematicaNotebookTwoLoop}. 
The colour-flow basis exposes a singnature-like symmetry where parity of the colour representation under $2\leftrightarrow 3$ permutation stands in one-to-one correspondence with the kinematic one. This suggests an interpretation of the spacelike-collinear limit in terms of effective degrees of freedom that adhere to this symmetry, namely Reggeons~\cite{Lipatov1976,Gribov1968,Regge1960,Regge1959,BalitskyLipatov1978,KuraevLipatovFadin1977,Gao:2024qsg,Moult:2022lfy,Falcioni:2021dgr,Byrne:2025phh,Falcioni:2021buo,Caron-Huot:2017fxr,Abreu:2024xoh,Caola:2021izf,Buccioni:2024gzo,DelDuca:2019tur,Caron-Huot:2013fea}. This is left for future exploration. 
Finally, we use the colour-flow basis to compare the $x$- and $y$-dependent HR  of the asymptotic amplitude ${\cal A}^{(2)}_{5}$ we computed, component by component, to the expected result~\cite{Dixon:2019lnw,Henn:2024qjq,Buccioni:2026mfg}, namely the right-hand side of Eq.~(\ref{Sp2}) where only the last term, 
${\mathbf{Sp}}^{(2)}
{\cal A}^{(0)}_{4}$, is relevant and where the kinematic dependence on $x$ and~$y$ is fully contained 
in~${\mathbf{\Delta}}_{\vec{\lambda}}^{(2)}$ of Eq.~(\ref{Delta2}). We find perfect agreement~\cite{MathematicaNotebookTwoLoop},  conforming our expectation that the HR alone is responsible for kinematically-dependent SCFV.

\section{Discussion}

We have shown by explicit computation   that the HR contribution to the asymptotic 5-point amplitude  $\mathcal{A}_5^{(2)}$ in the spacelike-collinear limit
captures the full SCFV kinematic dependence of the splitting amplitude ${\mathbf{Sp}}^{(2)}$ on the non-collinear partons in Eq.~(\ref{Delta2})~\cite{Dixon:2019lnw,Henn:2024qjq,Buccioni:2026mfg}. 
This distinctive role of the HR was predicted based on rather general considerations, namely the fact that the HR appears only in the spacelike collinear limit, while the same facets regions appear in both the timelike and spacelike limits. 

In momentum space the HR is characterised by soft interaction between the collinear particles and the rest of the process, as depicted in Figs.~\ref{cdep_graph_subtopology_Glauber_region}--\ref{possibly_hidden_singular_graph_dot_s}, each containing two soft lines attached to $p_1$ and $p_4$ (or $p_5$) respectively. It follows that one may apply eikonal approximations on partons $1$, $4$ ($5$) and the corresponding collinear propagators, equivalently replacing them by Wilson lines.
This explains, a posteriori, why the Wilson-line based calculation in Ref.~\cite{Dixon:2019lnw} agrees with the amplitude-based computation of ${\mathbf{Sp}}^{(2)}$~\cite{Henn:2024qjq,Buccioni:2026mfg}, and why the result in Eq.~(\ref{Delta2}) depends solely on the directions of the non-collinear partons rather than on their momenta and partonic species. 
Finally, the soft nature of the interaction and the connection with Wilson lines also explains the universality of the kinematic SCFV between the QCD and $\mathcal{N}=4$ sYM splitting amplitudes observed in~\cite{Buccioni:2026mfg}.

Our work also sheds light on recent work in the context of jet cross sections~\cite{Becher:2024kmk,Becher:2025igg,Becher:2026kbr,Dasgupta:2025cgl,Banfi:2025mra}, which are sensitive to Glauber effects.
The region responsible for restoration of collinear factorisation in this context (Fig.~3 in Ref.~\cite{Becher:2024kmk}) is a HR, precisely the real-emission counterpart to the one responsible for the SCFV we found here! This real-virtual connection of regions is expected to be general and can be useful in studying jet observables.

The methodology we develop to systematically identify HRs -- the HRF -- opens the way to study a broad range of kinematic limits of multi-leg amplitudes, including multi-collinear~\cite{Cieri:2024ytf,Guan:2024hlf,Duhr:2025cye,Duhr:2025lyg} and multi-Regge limits~\cite{Gao:2024qsg,Moult:2022lfy,Falcioni:2021dgr,Byrne:2025phh,Falcioni:2021buo,Caron-Huot:2017fxr,Abreu:2024xoh,Caola:2021izf,Buccioni:2024gzo,DelDuca:2019tur,Caron-Huot:2013fea} where Glauber effects are essential. 
This can also lead to more complete effective-theory descriptions of these limits and better understanding of the connections between Glauber soft-collinear effective theory and the Reggeon-based approach.

From a general perspective our results suggest that HRs provide a direct link between the Landau singularities governing asymptotic expansions~\cite{Jantzen:2012mw,Gardi:2024axt} and the crossing analyticity of scattering amplitudes~\cite{Bros:1985gy,Caron-Huot:2023ikn}. Establishing this correspondence in general will be an interesting direction for future work.

\begin{acknowledgments}
{\em Acknowledgements ---} We thank Hanyu Fang, Sebastian Jaskiewicz, and Kai Yan for discussions at early stages of this work. 
We also thank Franz Herzog and Stephen Jones for collaboration on related topics. WC is supported by the Natural Science Foundation of China (NSFC) under the contract number 11975200.
EG is supported by the STFC Consolidated Grant \emph{Particle Physics at the Higgs Centre}.
RM was supported by the Outstanding PhD Students Overseas Study Support Program of University of Science and Technology of China.
YM was supported by the Swiss National Science Foundation under the grant number 10001706.
YZ is supported by NSFC through Grant numbers 12575078 and 12247103.
ZZ was supported by the China Scholarship Council PhD programme and the European
Research Council (ERC) under the European Union’s Horizon Europe research and innovation
program grant agreement 101163627 (ERC Starting Grant ``AmpBoot''). 
For the purpose of open access, the authors have applied a Creative Commons Attribution (CC BY) licence to any Author Accepted Manuscript version arising from this submission.
\\
{\emph{Note Added:}} near the completion of this paper, we have been made aware of a related work by Barcaro, Gao, Gaunt, and Pathak~\cite{bggp}. Both  have been presented at the Parton Shower and Resummation conference in Manchester, 8-10 July~2026.
\end{acknowledgments}

\bibliography{biblio}
\bibliographystyle{apsrev4-1}

\section{End matter}

\noindent\textbf{A. Computing region integrals}

We compute the UT-basis integrals containing the scalar topologies discussed above, and include the relevant numerator directly in parameter space. For each such integral we compute the HR. The full computation will be discussed in~\cite{LongerPaper}. Here we proceed with an illustrative example, again considering $I_{\mathcal{G}_{\bullet\bullet}}^{\text{HR}}$ of Eq.~(\ref{IsixProp}) at leading power.  

The method we use is a variation on the dissection proposed in~\cite{Gardi:2024axt}.
To this end we first separate the $x_4$ integration defining 
${\cal D}{\bf x}_{\hat{4}}
\equiv 
\Pi_{i\neq 4} 
\mathrm{d}x_i\delta(1-\mathcal{E}({\bf x}))$
and 
extend the $x_4$ support:
\begin{align}
\begin{split}
I_{\mathcal{G}_{\bullet\bullet}}^{\text{HR}}&=
\int_{-\infty}^{\infty}dx_4
\int_0^\infty{\cal D}{\bf x}_{\hat{4}}\,{\cal I}_{\mathcal{G}_{\bullet\bullet}}
-
\int_0^\infty{\cal D}{\bf x}\,
\left.{\cal I}_{\mathcal{G}_{\bullet\bullet}}\right\vert_{x_4\to -x_4}\,
\\
&=
\int_{-\infty}^{\infty}dx_4^\prime
\int_0^\infty{\cal D}{\bf x}_{\hat{4}}\,
\left.
{\cal I}_{\mathcal{G}_{\bullet\bullet}}
\right\vert_{x_4\to x_4^\prime+(z-1)x_3}
\\&-
\int_0^\infty{\cal D}{\bf x}\,
\left.{\cal I}_{\mathcal{G}_{\bullet\bullet}}\right\vert_{x_4\to -x_4}
\end{split}
\end{align}
where 
in the first line we added and subtracted the contribution from $x_4<0$ where no pinch exists, and in the second placed the singular hypersurface~$x_4=(z-1)x_3$ at the origin, $x_4^\prime=0$, so the HR would appear as a facet region in the new coordinates for both $x_4^\prime<0$ and $x_4^\prime>0$. 
Using the Newton polytope in the new parameters the corresponding scaling is:
\begin{equation}
\label{vHR-shifted}
\vec{v}_{\HR}^{\,\prime}
\!=\!(v_{1},v_{2},v_{3},v_{4}^\prime,v_{5},v_{6};\!1)\!=\!
(-2, -1, -2, -1, -2, -1;\!1)
\end{equation}
which, as expected, coincides with (\ref{vHR_sol}) apart from the scaling of the shifted parameter $x_4^\prime\to x_4^\prime\delta^{v_{4}^\prime}$, which incorporates the cancellation. The obtained parametric integrals can be evaluated using the methods developed in Refs.~\cite{Chen:2019mqc,Chen:2019fzm,Chen:2020wsh,Chen:2025kdy,Chen:2023hmk}, which are implemented in the package {\tt AmpRed}~\cite{Chen:2024xwt,Chen:2025paq}. 

An important property of the region expansion at hand is that both the facet and hidden regions scale as~$\delta^{-4\epsilon}$. This means that these regions can in principle mix. The degeneracy of regions may give rise to divergences not regulated by the spacetime dimensions~\cite{Chen:2024mbk}. From the momentum-space perspective such mixing corresponds to propagators 1 and 5 in Fig.~\ref{cdep_graph_subtopology_collinear_region} carrying ${\cal O}(\delta)$ ``+'' components. In terms of the integrals, such mixing is realised as rapidity divergences of the separate facet and hidden regions, which in turn require an additional regulator. The dependence on the regulator must cancel in each integral between the facet and hidden regions (such divergences are not present in the seed integral $I_{\mathcal{G}_{\bullet\bullet}}$ we analysed above). 
Importantly, these divergences do not modify the conclusions we reached regarding the absence of kinematic dependence on $x$ and $y$ in the facet regions. 

To show this we regularise the rapidity divergence by raising the momentum space propagator 2 in Fig.~\ref{figure-cdep_graph_subtopology_representative_regions} to power~$a$ before the region expansion (note that the scaling of this propagator is different in the facet and hidden regions). Carrying this through the parameter space computation, both the facet region and the HR acquire dependence on $a$.
Since the LP polynomial of the facet region does not involve $y$ dependence at the outset, introducing an analytic regulator cannot change that. The situation could be more subtle for the $x$ dependence, but we find that for all integrals that require regularisation, the $x$ dependence can be removed from the LP polynomial by rescaling a suitable integration parameter which is not affected by the regulator. 
\\

\noindent\textbf{B. Colour bases for the spacelike-collinear limit of  $gg\to ggg$ amplitude }

To examine the asymptotic behaviour of the $2\to 3$ gluon amplitude, we decompose $\mathcal{A}_5^{(2)}$ in the standard trace basis~\cite{Edison:2011ta,Caron-Huot:2020vlo} and also a $t$-channel colour-flow basis~\cite{Abreu:2024xoh} used in the multi-Regge limit,
\begin{align}
\label{eq:AmpColour}
\mathcal{A}_5^{(2)}=\sum_{i=1}^{22}C_{\text{Tr,5}}^{i} A_{\text{Tr},5}^{(2),i}=\sum_{i=1}^{22}C_t^{[a_i,b_i]} A_{t,5}^{(2),i},
\end{align}
where the subscripts $\text{Tr}$ and $t$ on the colour-stripped amplitudes $A$ and colour tensors $C$ label the trace basis and $t$-channel colour-flow basis, respectively. The superscripts of $C_t^{[a_i,b_i]}$, $a_i$ and $b_i$, stand for representations of $SU(N_c)$ associated with the decomposition of $8_1\otimes 8_5$ and $8_2\otimes 8_3$ respectively, where the subscript of $8_i$ labels the corresponding external gluon. The components of the two bases are related by a $22\times 22$ rotation matrix $\mathbf{M}_{\text{Tr}\to t}$~\cite{Abreu:2024xoh},
\begin{align}
  \label{ColourRot}A_{t,5}^{(2),i}=\sum_{j=1}^{22}\left(\mathbf{M}_{\text{Tr}\to t}\right)_{ij} A_{\text{Tr},5}^{(2),j}. 
\end{align}
The matrix $\mathbf{M}_{\text{Tr}\to t}$ can be found in {\tt ColourRotation.m} in~\cite{MathematicaNotebookTwoLoop}. The colour-flow basis $\{C_t^{[a_i,b_i]}\}$ is natural in the context of $2,3$-collinear expansion because it diagonalises the $t$-channel colour matrices $\mathbf{T}_{t_1}^2\equiv(\mathbf{T}_1+\mathbf{T}_5)^2$ and $\mathbf{T}_{t_2}^2\equiv(\mathbf{T}_2+\mathbf{T}_3)^2$,
\begin{align}
\label{DiagT}
\begin{split}
   \mathbf{T}_{t_j}^2 \left[C_t^{[a_i,b_i]}A_{t,5}^{(2),i}\right]=f_j^i(N_c)C_t^{[a_i,b_i]}A_{t,5}^{(2),i},
   \\ i=1,\ldots,22,\quad j=1,2,
   \end{split}
\end{align}
where $f_j^i$ is a function of $N_c$ (see App. B.4 in Ref.~\cite{Abreu:2024xoh}).
In this basis, $x$ and $y$ dependence is present in~$12$ out of the~$22$ components in Eq.~(\ref{eq:AmpColour}), with $a_i\notin \{8^s\}$ or $b_i\notin\{8^s, 10\pm\overline{10}\}$.
Given that $2\leftrightarrow 3$ permutation manifests itself as  $\big\{z\to 1-z,\,\, y\to  1/y\big\}$, Bose symmetry links parity in $y\to  1/y$ with parity of the colour representation under $2\leftrightarrow 3$. We find that in the $t$-channel basis, odd functions under $y\to  1/y$ are associated with colour-even representations while even functions under $y\to  1/y$ with colour-odd ones~\cite{MathematicaNotebookTwoLoop}. This is closely related to the signature symmetry in the context of the Regge limit. 

To compare the $x$ and $y$ dependence of the HR to that of the splitting amplitude, we also consider $\mathbf{Sp}^{(2)}{\cal A}_4^{(0)}$ in Eq.~(\ref{Sp2}) in the same colour bases. Nontrivial dependence on $x$ and $y$ appears only through
\begin{align}
\label{xyDepTerm}
\begin{split}
&\mathbf{\Delta}^{(2)}\mathbf{T}^c_{c_2c_3}\sum_{i=1}^{6}C_{\text{Tr},4}^i(c_1,c_4, c_5, c)\,\,A_{\text{Tr},4}^{(0),i}
\\&=\mathbf{\Delta}^{(2)}\sum_{i=1}^{22}C_{\text{Tr},5}^i(c_1,c_2,c_3,c_4,c_5)\tilde{ A}_{\text{Tr},5}^{(0),i},
\end{split}
\end{align}
where $\mathbf{T}^c_{a_2a_3}$ comes from tree-level $\mathbf{R}_{\vec{\lambda}}$ in Eq.~(\ref{SpExp}) with superscript~$c$ representing the colour generator of the parent particle and subscript~$c_i$ representing that of the external particle~$i$. The tree-level 4-point amplitude $ A_{\text{Tr},4}^{(0),i}$ is decomposed in a trace basis $\{C_{\text{Tr},4}^i\}$ with colour indices $\{c_1,c_4,c_5,c\}$. Therefore, $\mathbf{T}^c_{c_2c_3}$ promotes the 4-point trace basis $C_{\text{Tr},4}^i$ to the 5-point trace basis $C_{\text{Tr},5}^i$ introduced in Eq.~(\ref{eq:AmpColour}) with indices $\{c_1,c_2,c_3,c_4,c_5\}$, which is described by the second line in Eq.~(\ref{xyDepTerm}). We also decompose Eq.~(\ref{xyDepTerm}) in the $t$-channel colour-flow basis,
\begin{align}
    \label{DeltaTrToT}\mathbf{\Delta}^{(2)}\sum_{i=1}^{22}C_{\text{Tr},5}^i\tilde{A}_{\text{Tr},5}^{(0),i}=\mathbf{\Delta}^{(2)}\sum_{i=1}^{22}C_{t}^{[a_i,b_i]}\tilde{A}_{t,5}^{(0),i}.
\end{align}
Note that $\tilde{A}_{\text{Tr},5}^{(0),i}$ and $\tilde{ A}_{t,5}^{(0),i}$ are related by the same rotation matrix appearing in Eq.~(\ref{ColourRot}).
To compute Eq.~(\ref{DeltaTrToT}), we express the colour-tripole structure $f^{abc}\mathbf{T}_1^a\mathbf{T}_3^b\mathbf{T}_I^d$ in $\mathbf{\Delta}^{(2)}$ of Eq.~(\ref{Delta2}) in terms of $\mathbf{T}_{t_1}^2$ or $\mathbf{T}_{t_2}^2$ and the following four operators, 
\begin{align}
    \begin{split}
       \mathbf{T}_{++}\equiv\,\,&(\mathbf{T}_1+\mathbf{T}_5)\cdot (\mathbf{T}_2+\mathbf{T}_3),
       \\\mathbf{T}_{+-}\equiv\,\,&(\mathbf{T}_1+\mathbf{T}_5)\cdot (\mathbf{T}_2-\mathbf{T}_3),
       \\\mathbf{T}_{-+}\equiv\,\,&(\mathbf{T}_1-\mathbf{T}_5)\cdot (\mathbf{T}_2+\mathbf{T}_3),
       \\\mathbf{T}_{--}\equiv\,\,&(\mathbf{T}_1-\mathbf{T}_5)\cdot (\mathbf{T}_2-\mathbf{T}_3).
    \end{split}
\end{align}
The  action of these colour operators on the $t$-channel colour-flow basis is already known~\cite{Abreu:2024xoh},
\begin{align}
\begin{split}
   \mathbf{T}_{\alpha\beta} \left[C_{t}^{[a_i,b_i]}\tilde{A}_{t,5}^{(0),i}\right]=C_{t}^{[a_i,b_i]}\sum_{j=1}^{22}\left(\mathbf{M}_{\alpha\beta}\right)_{ij} \tilde{A}_{t,5}^{(0),j},
   \\\alpha,\beta \in \{+,-\}.
   \end{split}
\end{align}
The transformation matrices $\{\mathbf{M}_{\alpha\beta}\}$ can also be found in~\cite{MathematicaNotebookTwoLoop}, and are stored in {\tt ColOpTPP.m}, {\tt ColOpTMP.m}, {\tt ColOpTPM.m} and {\tt ColOpTMM.m}. Now, we have both sides of Eq.~(\ref{Sp2}) in the $t$-channel colour-flow basis. Performing the comparison we confirmed that $x$ and $y$ dependence of the splitting amplitude $\mathbf{Sp}^{(2)}$ originates exclusively from the HR of the spacelike-collinear asymptotic expansion of the 5-point amplitude $\mathcal{A}_5^{(2)}$.
\end{document}